\documentclass[conference]{IEEEtran}

\ifCLASSINFOpdf
  \usepackage[pdftex]{graphicx}
  \graphicspath{{figures/}}
\else
\fi

\usepackage{amsmath}
\usepackage{amssymb}
\usepackage{flushend}

\begin{document}
%
\title{Prediction of the Initial Magnetic Field Properties of the CMEs Based on Simulated L1 Dynamics}


\author{\IEEEauthorblockN{Djalil Sawadogo}
\IEEEauthorblockA{Computer Science Department\\
Georgia State University\\
Atlanta, GA, USA\\
Email: asawadogo2@student.gsu.edu}
\and
\IEEEauthorblockN{Dustin Kempton}
\IEEEauthorblockA{Computer Science Department\\
Georgia State University\\
Atlanta, GA, USA\\
Email: dkempton1@gsu.edu}
\and
\IEEEauthorblockN{Viacheslav M Sadykov}
\IEEEauthorblockA{Physics \& Astronomy Department\\
Georgia State University\\
Atlanta, GA, USA\\
Email: vsadykov@gsu.edu}
\and
\IEEEauthorblockN{Elena Provornikova}
\IEEEauthorblockA{The Johns Hopkins University Applied Physics Laboratory\\
Laurel, MD, USA\\
Email: elena.provornikova@jhuapl.edu}
}

\maketitle

\begin{abstract}
Coronal Mass Ejections (CMEs) are among the primary drivers of space weather impacts, but physics-based forecasting of their terrestrial effects depends on a magnetic field structure that is not directly observable routinely and must be prescribed at initialization. Analytical flux rope models such as the Gibson-Low (GL) formulation reduce this structure to a small set of physical parameters: field strength (\verb|Bpar|), topology/morphology (\verb|TOPMORPH|), flux-rope orientation, and chirality. Nevertheless, constraining these parameters for a given event still typically relies on iterative, computationally expensive forward modeling. This work poses the inverse problem instead: given the CME's in-situ magnetic field and plasma time series at L1, its kinematic properties, and background solar wind conditions, can its initial GL flux-rope parameters be recovered directly? We train and compare a range of predictive models (from time-series-forest and recurrent (LSTM-based) architectures to Transformer-based models) on a grid of over 23,000 GAMERA-GL MHD simulations spanning the GL parameter space, mapping L1 time series and background solar wind spherical harmonic context to the GL parameters. The Transformer architecture was selected as the final model because it outperformed the other approaches on every target, while the comparison across architectures showed that chirality cannot be predicted above chance level from L1 dynamics, which led the final model to focus on the three continuous flux-rope parameters. Prediction accuracy proved strongly dependent on CME magnetic imprints at L1, reaching $R^2$ values of 0.967, 0.860, and 0.833 for the CME \verb|TOPMORPH|, \verb|Bpar|, and orientation properties, correspondingly for the CMEs delivering peak $|B| > 20$\,nT impact. These results show L1 in-situ dynamics carry substantial, recoverable information about a CME's initial magnetic structure in case of CMEs with strong magnetic field, offering a fast, data-driven complement to trial-and-error forward-model fitting.
\end{abstract}

\begin{IEEEkeywords}
coronal mass ejections, space weather, inverse problems, deep learning, MHD simulations
\end{IEEEkeywords}

%
\IEEEpeerreviewmaketitle

\section{Introduction}

    Understanding the impact of solar coronal mass ejections (CMEs) on the terrestrial environment is among the primary challenges in space weather. CMEs can cause a variety of effects, including geomagnetically induced currents in power grids and power lines and increased atmospheric drag on low-Earth-orbit satellites due to enhanced thermospheric density. The geoeffectiveness of the CME is determined by various properties, with the arrival time and $B_z$ profile being of the most operational interest. The $B_z$ dynamics is closely related to the geomagnetic disturbance typically quantified by geomagnetic indices such as Kp \cite{Elliott:2013:kp_to_sw_relation} and Dst \cite{Gonzalez:2005:peakDst_to_Bz,Kane:2010:Dst_SC23}. The knowledge of the expected $B_z$ is therefore essential, and the related prediction challenge is typically approached by with physics-based modeling of the CME propagation \cite{Shiota:2016:MHD_SUSANOO,Torok:2018:MHD_suntoearth,Jin:2017:EEGGLmodel}.

    One of the main challenges of such modeling is the initialization of the CME close to the Sun. While the kinematics and geometry of the CME during its propagation near the Sun can be reconstructed from the white-light coronagraph observations (by the coronagraphs such as LASCO/SOHO \cite{Brueckner1995SolarPhysics} or STEREO/SECCHI \cite{Kaiser2008STEREO}), the question of prescribing the internal magnetic field of the CME remains essentially unsolved due to a lack of observations that could provide information on the magnetic field. Some approaches try to constrain the CME magnetic properties based on the magnetic configuration in the host active regions (ARs) and regions therein such as line polarity inversion lines \cite{Borovikov:2017:EEGGL,Jin:2017:EEGGLmodel}. Nevertheless, the question remains of how the AR magnetic field properties map into the magnetic field properties of the CMEs during their eruption and evolution in the solar corona. One way to improve understanding of this problem is to perform a detailed modeling of a number of the historic CME events and compare their properties to host ARs.

    However, such an approach is computationally costly. One ideally would like the modeled CME to (1) be constrained in its geometry by coronagraph observations, ideally stereoscopic, (2) propagate in the background solar wind corresponding to the conditions of a particular date, and (3) match its profiles at in-situ observational points, i.e., the Lagrange 1 point. The latter is subject to not only CME geometry but magnetic field properties as well, which for the Gibson-Low flux rope configuration \cite{Malanushenko:2020:CoNNs,Provornikova2024GAMERAGL} are:
    \begin{itemize}
        \item The maximum magnitude of the magnetic field in a CME (hereafter \verb|Bpar|);
        \item The topology and morphology of the flux rope \cite{Malanushenko:2020:CoNNs,Provornikova2024GAMERAGL} (hereafter \verb|TOPMORPH|);
        \item The orientation of the flux rope (hereafter \verb|orientation|);
        \item The chirality of the field (hereafter \verb|CHIRAL|).
    \end{itemize}
    Iterating over the grid of these parameters in order for the CME to match the observed dynamics at L1 can quickly result in a large number of simulations and become costly. While chirality is a binary parameter (can be either right-handed or left-handed only), the others can take continuous values. Therefore, the possibility of providing a good initial estimate for the parameters of the CME near the Sun is an important problem.

    There is a possibility to optimize CME model initiation using large grids of simulation data. In this work, we use a simulation grid that has been generated with the GAMERA-Helio MHD model of the inner heliosphere \cite{Zhang:2019:GAMERA,Merkin:2016:GAMERA_a,Merkin:2016:GAMERA_b}. The model is coupled with a Gibson-Low CME flux rope \cite{Provornikova2024GAMERAGL, Gibson:1998:GL}, and uses the Wang--Sheeley--Arge (WSA) model\cite{Arge:2000:WSA,Arge:2004:WSA} to provide the boundary condition of the ambient solar wind at the 21.5\,R$_{Sun}$. The grid contains 23,058 simulation runs, with three background solar wind boundary conditions, 22 unique CME launch locations, and covering various CME properties (fast/slow CMEs, wide/narrow CMEs, CMEs with strong or weak magnetic fields, etc). From this grid, we sample more than 7 million unique time series of solar wind dynamics at the L1 point. The simulation grid properties and further downselection of L1 points are discussed below.

    The primary science problem addressed in this paper is to investigate whether, and how strongly, one can constrain the CME magnetic field properties near the Sun, based on the known CME geometry (speed and width properties), launch location with respect to L1, background solar wind conditions, and the CME dynamics observed in situ at the L1 point. The motivation is to later utilize the developed models to constrain the magnetic field properties of some historical CME events, and therefore assist the time-consuming forward modeling exercise to be performed in scale. The paper is organized as follows: Section~\ref{sec:methodology} describes the methodology aspects of this work, including some dataset details and model selection and training details. Section~\ref{sec:results} presents the results of this work, followed by conclusion statement in Section~\ref{sec:conclusion}.

\section{Methodology}\label{sec:methodology}

    In this section, we describe the data used to pose the inverse problem, the target parametrization, the L1 input representation provided to the models, the progression of models evaluated before the final architecture, the final model architecture, and the training and evaluation setup used to assess its performance.

\subsection{Modeling Suite and L1 Timeseries}

    The simulation grid was produced with a coupled model suite. The Wang--Sheeley--Arge (WSA) model with ADAPT solar magnetograms \cite{Arge:2000:WSA,Arge:2004:WSA} approximates the coronal magnetic field reconstructed from photospheric magnetic field observations. The WSA-ADAPT results for the radial magnetic field $B_r$ and radial solar wind speed $v_r$ at 21.5\,R$_{Sun}$ serve as the inner boundary conditions for the MHD GAMERA-Helio model of the inner heliosphere \cite{Zhang:2019:GAMERA,Merkin:2016:GAMERA_a,Merkin:2016:GAMERA_b,Provornikova2024GAMERAGL}. GAMERA then computes the steady background solar wind solution (in the frame of reference rotating with the Sun) from 0.1 to 1\,AU by running the MHD solver for about 200 hours (in physical time units) in order for the solution to relax. For the background condition, three Carrington Rotations were utilized: CR 2095, CR 2194, CR 2226 (rising, declining, and minimum phases of solar activity, respectively).
    
\begin{figure*}[!t]
        \centering
        \includegraphics[width=\textwidth]{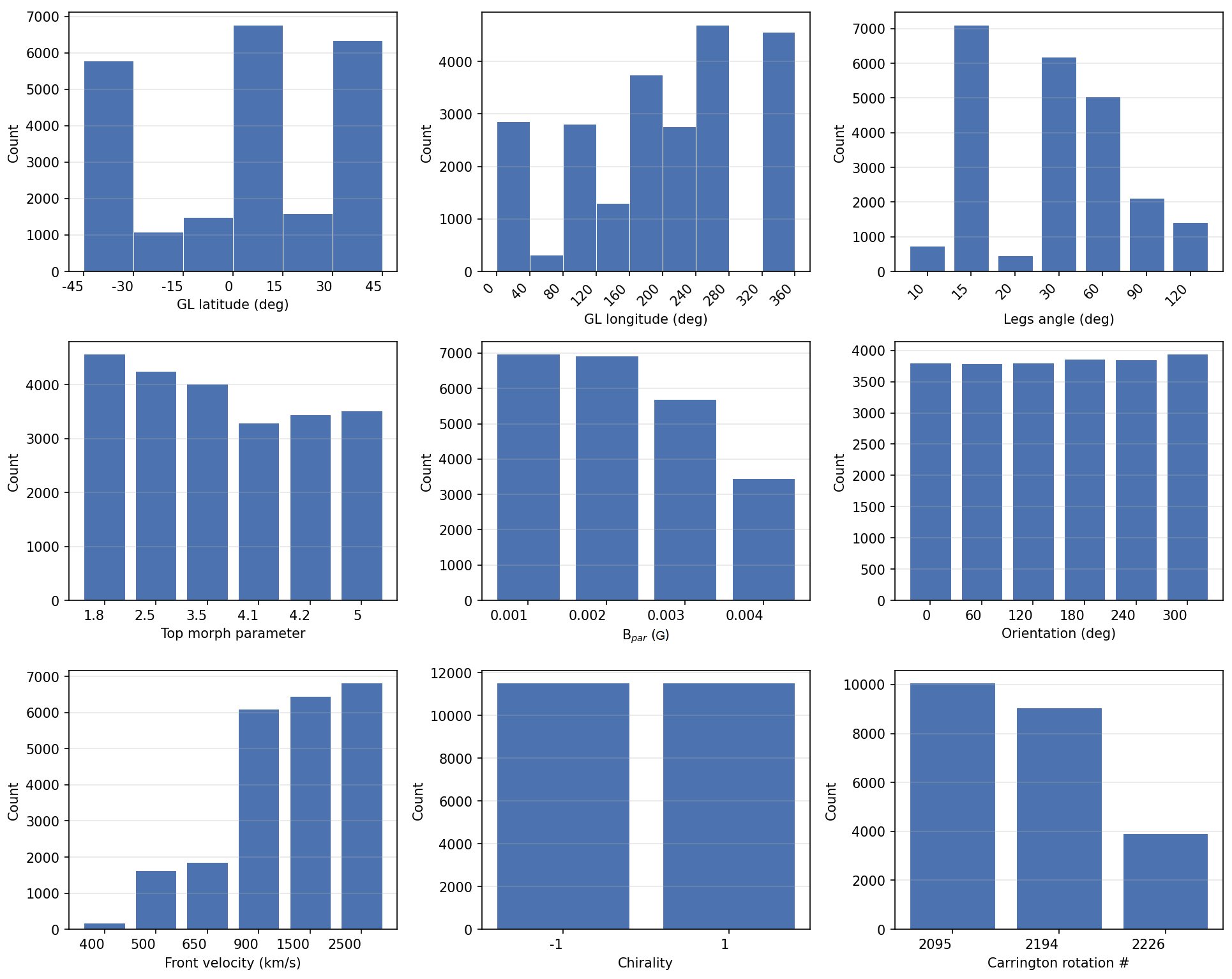}
        \caption{A histogram of the Gibson-Low CME parameters at the CME initiation.}
        \label{fig:cmeparameters}
    \end{figure*}

    CMEs are then injected through the inner boundary (at 21.5\,R$_{Sun}$ using the analytical Gibson-Low (GL) solution of a self-similarly expanding magnetic flux rope \cite{Gibson:1998:GL,Provornikova2024GAMERAGL}. We note here that, while the analytical solution was used to initialize the CME structure near the GAMERA inner boundary, the further evolution of the flux rope beyond 21.5\,R$_{Sun}$ has been controlled by the MHD equations and solved by GAMERA. The parameters of the injected CMEs are illustrated in Figure~\ref{fig:cmeparameters}.

\begin{figure*}[!t]
        \centering
        \includegraphics[width=\textwidth]{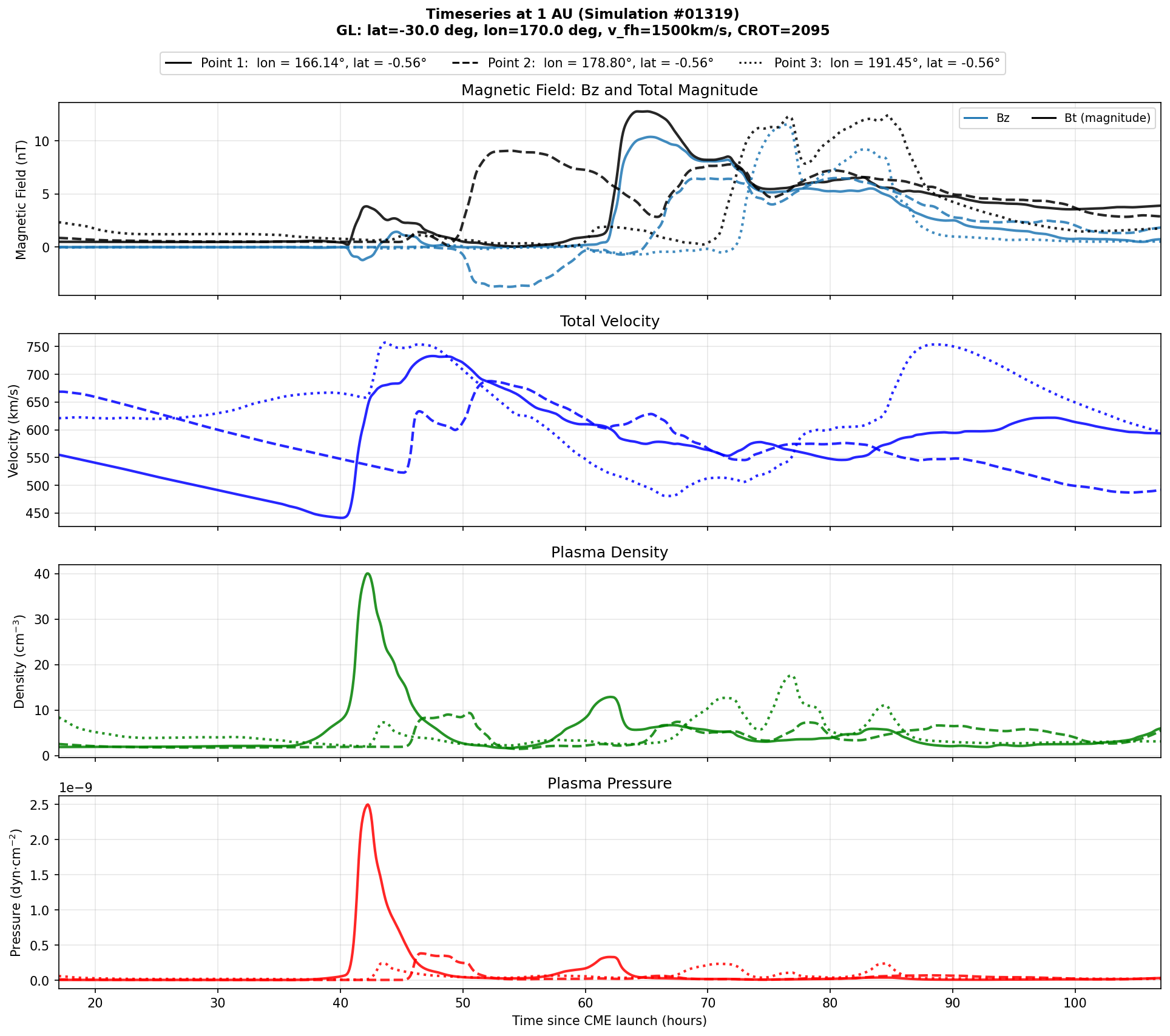}
        \caption{Examples of the CME timeseries at the 1\,AU shell sampled within the dataset creation effort.}
        \label{fig:L1dynamics}
    \end{figure*}

    The simulations were later saved as 1\,AU (Astronomical Unit) shells in time. Given that the 1\,AU and L1 are physically separated by just about 1\% of the distance from the Sun to Earth, here we will refer to the terms ``1\,AU shell'' and ``L1 shell'' interchangeably. The shells tracked the passage of the CME, up to 120\,hours from the CME launch time (for the case of the slowest CMEs). We then sample the CMEs at the locations that hypothetically can correspond to the L1 points. We sample the L1 shells with the step of $\sim$4$^\circ$ along the longitude modulated with the longitude-dependent latitudinal offset. Along latitude, the locations are drawn within $\sim$8$^\circ$ of the simulation equator, with a comparable angular stepping. This results in $\gtrsim$10.5 million time series in total. At each location, the magnetic field and velocity vectors are converted from the simulation frame to the Geocentric Solar Ecliptic (GSE) system, in which L1 measurements are commonly taken. We record time series of $B_x$, $B_y$, $B_z$, $|B|$, $v_x$, $v_y$, $v_z$, $|v|, $density $\rho$, and pressure $p$. The timeseries have been interpolated to the cadence of 5\,min and the total exact length of 90\,h, constituting 1081 time points per simulation, to mitigate variations of the time steps in individual simulations. At the same time, the beginning time moment starting from which the L1 series are recorded depends on the initial CME speed (being later for lower CMEs). The examples of the sampled timeseries for the selected simulation run are presented in Figure~\ref{fig:L1dynamics}. More information about the dataset was presented in \cite{Sadykov2026COSPAR} (available by request) and is described in the paper currently in preparation \cite{Sadykov:2026:MLdataset}.

\subsection{Downselection for Inverse Modeling}

    We further select the specific dataset points down for the inverse problem. It is understandable that the inversion for the magnetic properties of the CME is potentially possible, in principle, only if the CME (1) is impacting L1 location, and (2) has a significant magnetic imprint at L1. The first condition restricts the launch sites of the CME to be at least within 90 degrees (longitudinally) from the location of the L1 point. We note here that, given the spatial variability of the solar wind, we cannot distinguish the CME-related and background solar wind-related variability in case of weak to moderate CMEs (which was a subject of a separate investigation). To address the second condition, we consider only the timeseries of the CMEs that had the maximum magnetic field at L1 to exceed 5\,nT. The latter is motivated by the real solar wind conditions, with the background solar wind often experiencing the comparable magnitudes of magnetic field variations. After applying both conditions, the dataset contains $807{,}810$ L1 time series.

\subsection{Target Parametrization}
\label{sec:target-param}

    The Gibson-Low (GL) flux-rope model specifies a CME's initial magnetic structure with four parameters: Maximum field strength (\verb|Bpar|), topology/morphology (\verb|TOPMORPH|), flux-rope orientation (\verb|Orientation|), and chirality (\verb|CHIRAL|). Of these, \verb|CHIRAL| is a discrete, binary-valued quantity (right- or left-handed) rather than a continuous one. As detailed in Section~\ref{sec:model-selection}, \verb|CHIRAL| did not prove reliably learnable from L1 dynamics under any of the architectures evaluated in this work, so the final model treats it as an input context feature (Section~\ref{sec:input-rep}) rather than an output.

    The remaining three parameters (\verb|TOPMORPH|, \verb|Bpar|, and \verb|GL_Orientation|) are treated as continuous regression targets, and are the three outputs of the final model's regression heads (Section~\ref{sec:model-arch}). Although all three are continuous physical quantities, the initial physics-based simulations are computationally expensive, so the simulations were run for a limited set of chosen values of each parameter: four values of \verb|Bpar| ($0.001$ to $0.004$), six of \verb|TOPMORPH| ($1.75$ to $5.0$), and six of \verb|GL_Orientation| ($0$ to $300$ in steps of $60$). See Figure~\ref{fig:cmeparameters} for related distributions.

\subsection{Input Representation}
\label{sec:input-rep}

    The model is given two complementary views of each event: a raw time-series input and a scalar context vector, described below, each rescaled per-channel as described in Section~\ref{sec:preprocessing}.

    \subsubsection{Time-Series Input}

    The model receives ten in-situ channels measured at L1: the three magnetic field components ($B_x$, $B_y$, $B_z$), field magnitude $B_m$, plasma density, plasma pressure, the three velocity components ($v_x$, $v_y$, $v_z$), and velocity magnitude $V_m$. Each channel is sampled at $1209$ timesteps per event. After the per-channel scaling described below, these series are passed to the time-series branch described in Section~\ref{sec:model-arch} in their full $10 \times 1209$ form, with no pooling or hand-engineered feature extraction applied beforehand. Unlike the scalar branch's summary statistics, below, the model's time-series branch receives the rescaled series itself.

    \subsubsection{Scalar Context Vector}

    The remaining, non-time-series columns are concatenated into the scalar context vector fed to the scalar branch. This vector has three components. First, real spherical-harmonic coefficients of the inner-boundary radial magnetic field ($B_r$) and radial velocity ($v_r$), decomposed up to degree $L=5$: each field contributes a cosine and a sine coefficient for every $(l,m)$ pair with $0 \le m \le l \le L$, giving $(L{+}1)(L{+}2)/2 = 21$ coefficient pairs per field and $21 \times 2 \times 2 = 84$ coefficients in total. Second, a set of scalar CME and observational parameters: the CME source latitude and (L1-relative) longitude (\verb|GL_lat|, \verb|GL_lon|), the flux-rope leg angle (\verb|GL_legsang|), front height and velocity at front height (\verb|GL_frontheight|, \verb|GL_vel_fh|), the CME density (\verb|Den_CME|), the chirality of the field (\verb|GL_CHIRAL|, here used as an input feature rather than a predicted target, see Section~\ref{sec:target-param}), and the hypothetical spacecraft's own latitude and longitude at L1 (\verb|lat_L1|, \verb|lon_L1|). Third, for each of the ten time-series channels above, twelve summary statistics computed over the full series: mean, standard deviation, minimum, median, maximum, skewness, kurtosis, linear-trend slope, standard deviation of the first difference, lag-1 autocorrelation, zero-crossing count, root-mean-square, and the value at the first timestep — giving the model access to hand-engineered summaries of the same series it also receives in raw form via the time-series branch.

    \subsubsection{Preprocessing}
    \label{sec:preprocessing}

    Every scalar and time-series channel is independently rescaled based on its empirical distribution in the training split: channels that are strongly right-skewed and non-negative (e.g., plasma density, pressure, and field magnitude, and their first-timestep scalar counterparts) are log-transformed and then standardized; signed channels with heavy tails that cross zero (e.g., $B_y$, $B_z$, $v_y$, $v_z$, and their first-timestep counterparts) are transformed with the inverse hyperbolic sine before standardization; every other channel is standardized directly (zero mean, unit variance) without a nonlinear transform.

\subsection{Model Selection}
\label{sec:model-selection}

    Before arriving at the final architecture (Section~\ref{sec:model-arch}), we evaluated a progression of modeling approaches, summarized here to guide future work on this problem. We first establish classical statistical baselines (based on statistical features of considered timeseries, Section~\ref{sec:stat-baselines}), then multi-task deep learning models that predict \verb|CHIRAL| jointly with the three continuous targets (Section~\ref{sec:mtl-baselines}). The resulting evidence that \verb|CHIRAL| is not reliably learnable under any of these architectures (Section~\ref{sec:chiral-pred}) motivates the final model, which instead takes \verb|CHIRAL| as an input feature (Section~\ref{sec:chiral-input}).

    \subsubsection{Statistical Baselines}
    \label{sec:stat-baselines}

    As an initial baseline, independent of any deep architecture choice, we trained two classical tree-based ensembles --- a random forest (RF) \cite{Breiman:2001:RF} and a gradient-boosted tree ensemble (XGBoost) \cite{Chen:2016:XGBoost}. Here, we trained one model per target, including \verb|CHIRAL|, using only the scalar context vector (Section~\ref{sec:input-rep}), excluding \verb|CHIRAL| and with no raw time series. We additionally evaluated Slim-TSF \cite{Ji:2024:SlimTSF}, a sliding-window multivariate time-series-forest method that instead derives its own fixed-size feature set directly from the raw L1 series (per-interval mean, standard deviation, and slope, pooled across several window scales) combined with bootstrapped feature selection, again feeding an RF model. Together, these establish how much of the problem a non-sequential (besides the Slim-TSF approach), statistical feature-engineered approach can capture before any recurrent or attention-based model is introduced.

    \subsubsection{Multi-Task Deep Learning Baselines}
    \label{sec:mtl-baselines}

    We next evaluated deep multi-task learning (MTL) architectures \cite{Caruana:1997:MTL} that predict \verb|CHIRAL| jointly with the three continuous targets as a fourth output, to test whether a shared deep representation could extract a chirality signal that the statistical baselines could not (as we will see further), along with other parameters. MTL-DNN is a fully-connected network over the scalar context vector alone, with no time-series branch. BiLSTM-MTL and Transformer-MTL share the dual-branch design of Section~\ref{sec:model-arch}, but with a bidirectional LSTM \cite{Hochreiter:1997:LSTM,Graves:2005:BiLSTM} or Transformer encoder, respectively, as the time-series branch, and four output heads instead of three. In all three MTL models, \verb|CHIRAL| is excluded from the scalar context vector and is initially a quantity to predict.

    \subsubsection{Chirality Predictability Across Architectures}
    \label{sec:chiral-pred}

    Table~\ref{tab:chiral} reports \verb|CHIRAL| classification performance for every architecture evaluated above. No architecture (statistical, recurrent, or attention-based) exceeds the majority-class baseline's accuracy, and every Matthews correlation coefficient (MCC) is at most $0.02$, indicating chance-level discrimination between left- and right-handed flux ropes regardless of model class. This consistent null result, rather than any single model's limitation, is what motivates removing \verb|CHIRAL| as a predicted target in the final model (Section~\ref{sec:chiral-input}). We note here that, given that \verb|CHIRAL| is a binary parameter (can be either left-handed or right-handed), the absence of its prediction increases the complexity of the follow-up forward modeling effort twice at maximum (and therefore represents a reasonable sacrifice for the further enhancement of no-chirality model accuracies).

\begin{table}[!t]
\renewcommand{\arraystretch}{1.2}
\caption{Test-Set Performance of the Evaluated Architectures on Chirality (\texttt{CHIRAL})}
\label{tab:chiral}
\centering
\begin{tabular}{lccc}
\hline
Model & Accuracy & Balanced Accuracy & MCC \\
\hline
RF              & 0.5041 & 0.5059 & 0.012 \\
XGBoost         & 0.5040 & 0.5047 & 0.010 \\
TSF             & 0.4985 & 0.5017 & 0.004 \\
MTL-DNN         & 0.5063 & 0.5064 & 0.013 \\
BiLSTM-MTL      & 0.5065 & 0.5065 & 0.013 \\
Transformer-MTL & 0.5100 & 0.5095 & 0.019 \\
\hline
Majority-class baseline & 0.5111 & 0.5000 & 0.000 \\
\hline
\multicolumn{4}{p{0.93\columnwidth}}{\footnotesize
RF: random forest; TSF: Slim-TSF time-series forest; DNN: deep neural network;
BiLSTM: bidirectional long short-term memory network; MTL: multi-task learning.
The majority-class baseline always predicts $-1$, the larger class
($83{,}435$ of $163{,}246$ test samples). No architecture exceeds its
accuracy, and all Matthews correlation coefficients (MCC) are at most $0.02$,
indicating chance-level discrimination between left- and right-handed
flux ropes.}
\end{tabular}
\end{table}

    \subsubsection{Chirality as an Input Feature}
    \label{sec:chiral-input}

    Given this result, the final model instead treats \verb|GL_CHIRAL| as a known scalar input feature (Section~\ref{sec:input-rep}) rather than an output, and predicts only the three continuous targets. The comparison in Section~\ref{sec:comparison} (Table~\ref{tab:comparison}) supports this choice: for both the BiLSTM and Transformer encoders, the version that receives \verb|CHIRAL| as an input attains a higher mean $R^2$ than its MTL counterpart that predicts it jointly. For this comparison, we retained the two architectures with distinct time-series encoders, transformer-based and recurrent, that performed acceptably when \verb|CHIRAL| was a target, and re-trained each with \verb|CHIRAL| as an input and without the \verb|CHIRAL| output head. Comparing the resulting BiLSTM and Transformer models isolates the effect of the time-series encoder. We note that \verb|CHIRAL| is a modeling input of the GL flux rope rather than a quantity constrained by the L1 dynamics. In practical applications, it can be inferred from observations of the CME source region, for example from the chirality and axial field direction of the erupting filament \cite{Aparna:2020:filamentsBz,Mundra:2021:CMEprogenitors}. Establishing a reliable mapping from such observations to the GL chirality is left to future work. Alternatively, both values for \verb|CHIRAL| may be tested, with subsequent forward modeling identifying the better-fitting one through comparison with L1 measurements. The performance of the model when chirality is unavailable is indicated by the Transformer-MTL model of Table~\ref{tab:comparison}, which does not receive \verb|CHIRAL| as an input and attains a mean $R^2$ of $0.406$, compared with $0.458$ when \verb|CHIRAL| is given.

\subsection{Model Architecture}
\label{sec:model-arch}

\begin{figure*}[!t]
    \centering
    \includegraphics[width=\textwidth]{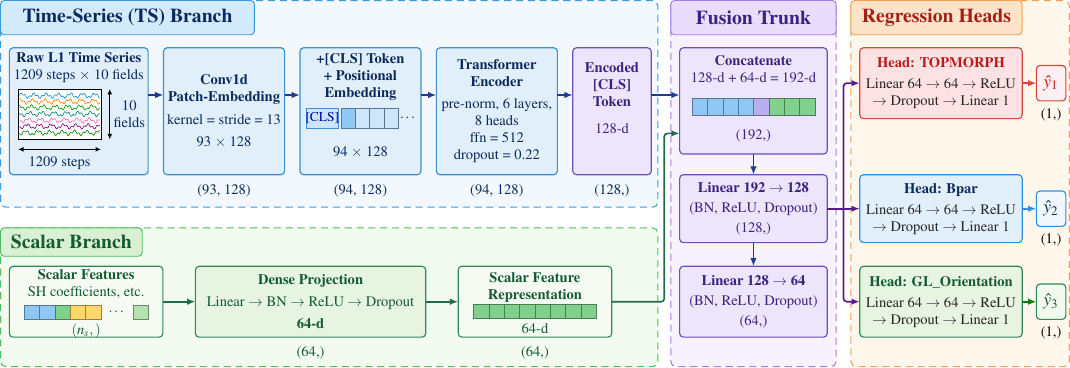}
    \caption{Dual-branch Transformer regression architecture. The time-series branch
    encodes the $10 \times 1209$ L1 series into a $128$-dimensional [CLS] representation,
    the scalar branch projects the scalar context vector to $64$ dimensions, the fusion
    trunk combines both, and three independent regression heads predict \texttt{TOPMORPH},
    \texttt{Bpar}, and \texttt{GL\_Orientation}.}
    \label{fig:architecture}
    \end{figure*}

    The final model used in this work (Fig.~\ref{fig:architecture}) is a dual-branch Transformer-based regressor operating on the L1 time series and the scalar context vector (spherical-harmonic coefficients together with the other scalar CME and observational parameters) described in Section~\ref{sec:input-rep}. A BiLSTM variant, sharing every other component but using a bidirectional LSTM in place of the Transformer encoder, was also carried forward as a direct point of comparison (Table~\ref{tab:comparison}). However, its performance has been worse in comparison with the Transformer-based architecture. Therefore, we describe the final Transformed architecture below.

    \subsubsection{Time-Series Branch}

    The raw multivariate L1 time series is first tokenized into fixed-length, non-overlapping patches along the time axis via a channel-mixing 1D convolution (every patch token is built jointly from all input channels), linearly embedding each patch into the model's working dimension $d_{\mathrm{model}}$ (Table~\ref{tab:hparams}). A learnable classification ([CLS]) token is prepended to the resulting patch-embedding sequence, a learned positional embedding is added, and the sequence is passed through a pre-norm multi-head self-attention Transformer encoder \cite{Vaswani:2017:attention}, which lets the model directly weigh the relevance of every patch to every other patch when forming its representation of the series, rather than propagating information step-by-step as in a recurrent architecture. The encoder's output at the [CLS] token position is taken as the branch's output: a single $d_{\mathrm{model}}$-dimensional vector per sample, summarizing the entire input series.

    \subsubsection{Scalar Branch}

    The scalar context vector is passed through a single fully-connected layer (Linear $\to$ BatchNorm $\to$ ReLU $\to$ Dropout) that projects it to a $64$-dimensional vector.

    \subsubsection{Fusion Trunk}

    The time-series branch's $128$-dimensional output and the scalar branch's $64$-dimensional output are concatenated into a single $192$-dimensional vector, which is the input to a shared fully-connected trunk of two further (Linear $\to$ BatchNorm $\to$ ReLU $\to$ Dropout) layers, of width $128$ and then $64$. The trunk's output is therefore a single $64$-dimensional vector per sample, shared identically across every target.

    \subsubsection{Regression Heads}

    Each of the three continuous targets (\verb|TOPMORPH|, \verb|Bpar|, \verb|GL_Orientation|) is produced by its own independent regression head, with no weight sharing between heads. Each head takes the shared trunk's $64$-dimensional output as its input and is itself a small two-layer MLP, Linear($64\to64$) $\to$ ReLU $\to$ Dropout $\to$ Linear($64\to1$), producing a single scalar prediction. The three heads' scalar outputs are concatenated into the model's final $3$-dimensional output vector.

\subsection{Training and Evaluation Setup}
\label{sec:train-eval}

    \subsubsection{Split}

    The dataset is split into training and held-out test sets using a \emph{group split} by simulation. Effectively, every timeseries sample belonging to a given simulated event is assigned entirely to one split or the other. This prevents the model from ever being evaluated on a sample very close to what it trained on, which a fully random split would risk given multiple samples per event. The held-out test set comprises $20\%$ of simulations and is fixed once, so all architectures are evaluated on the same test set. The training simulations are further divided, again by simulation, into a fitting set and a validation set, with approximately $10\%$ of the training simulations assigned to validation. The validation set is used for early stopping, for selecting the best epoch, and as the objective of the hyperparameter search (Section~\ref{sec:hparam}). The test set is used only for the final evaluation. We note that the simulations that share most of their parameters but one (for example, otherwise identical CMEs of opposite chirality) may still be assigned to different splits. The effect of such configurations on the reported performance has not been investigated here and is left to future work.

    \subsubsection{Hyperparameter Tuning}
    \label{sec:hparam}

    The learning rate, weight decay, dropout rate, encoder depth, embedding width $d_{\mathrm{model}}$, and patch size were selected via an automated search. A Tree-structured Parzen Estimator (TPE) sampler \cite{Bergstra:2011:TPE}, as implemented in Optuna \cite{Akiba:2019:Optuna}, was used to minimize the validation loss over learning rate and weight decay (log-uniform over $[10^{-5}, 10^{-2}]$ and $[10^{-6}, 10^{-2}]$ respectively), dropout (uniform over $[0, 0.5]$), encoder depth ($\{2,3,4,6\}$ layers), $d_{\mathrm{model}} \in \{64, 128, 256\}$, and patch size $\in \{5, 8, 13\}$ timesteps. The resulting values, used throughout Sections~\ref{sec:model-arch} and~\ref{sec:train-eval}, are given in Table~\ref{tab:hparams}.

    \begin{table}[!t]
    \renewcommand{\arraystretch}{1.3}
    \caption{Tuned Hyperparameters}
    \label{tab:hparams}
    \centering
    \begin{tabular}{|c|c|c|c|c|c|}
    \hline
    LR & WD & Drop. & Layers & $d_{\mathrm{model}}$ & Patch \\
    \hline
    $1.010{\times}10^{-3}$ & $2.108{\times}10^{-6}$ & $0.0077$ & $6$ & $128$ & $13$ \\
    \hline
    \end{tabular}

    \vspace{2pt}
    \footnotesize LR: learning rate. WD: weight decay. Drop.: dropout rate. Patch: patch size (timesteps).
    \end{table}

    \subsubsection{Optimization}

\begin{figure}[!t]
    \centering
    \includegraphics[width=\columnwidth]{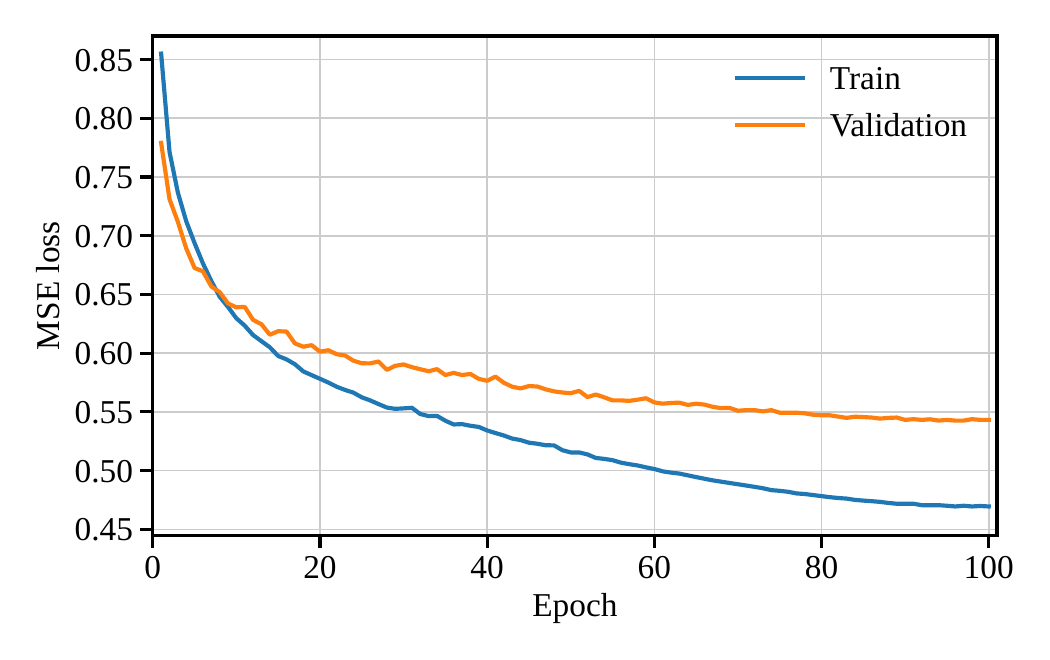}
    \caption{Training and validation loss (MSE summed over the three regression targets) of the final Transformer model per epoch.}
    \label{fig:loss}
    \end{figure}

    The model is trained with the AdamW optimizer (learning rate and weight decay per Table~\ref{tab:hparams}) under a plain cosine-annealing learning-rate schedule (no warmup phase), stepped once per epoch over a maximum of $100$ epochs, with early stopping on the validation loss (patience of $10$ epochs with no improvement), and the weights from the epoch with the lowest validation loss are retained. The loss is the mean squared error (MSE) summed over all three regression targets. Fig.~\ref{fig:loss} shows the training and validation loss over the course of training.

    Both losses continue to decrease slowly through the full $100$-epoch budget without ever triggering the patience-$10$ early-stopping criterion. This fixed epoch count was applied uniformly across every PyTorch-based architecture compared in this work, rather than tuned independently per model, to keep training budgets comparable; the pronounced flattening of both curves beyond approximately epoch $70$ suggests that additional epochs would yield only marginal further improvement.

    \subsubsection{Evaluation Metrics}

    All architectures are evaluated on the same held-out test set with the same metrics. Each continuous target is evaluated independently with the coefficient of determination ($R^2$), the root-mean-squared error (RMSE, in the target's own units), Pearson and Spearman correlation between true and predicted values, and the mean signed error (bias, predicted minus true); architectures are compared by $R^2$ averaged over the three targets. The Pearson correlation between the prediction errors of each pair of targets is also computed, to assess whether errors on one target accompany errors on another. For the models that predict \verb|CHIRAL|, the class prediction is the sign of the regression output, and classification performance is measured by accuracy, balanced accuracy (the mean of the per-class recall rates), and the Matthews correlation coefficient (MCC), and compared with a majority-class baseline that always predicts the more frequent class. Each test row (one L1 time series) is additionally assigned to one of four CME-severity bins according to the peak magnetic field magnitude $B_m$ observed at L1 in that row ($5$--$10$, $10$--$15$, $15$--$20$, and $>20$~nT), and all metrics are recomputed within each bin, relative to that bin's own target variance, to assess whether prediction accuracy depends on the strength of the driving CME. The bins are used for evaluation only: a single model is trained on the full training set, spanning all $B_m$ options, and its test-set predictions are then grouped by bin.

\section{Results}\label{sec:results}
\label{sec:results}

\subsection{Comparison of Architectures}
\label{sec:comparison}

    Table~\ref{tab:comparison} returns us to the comparison of the final model with the other architectures evaluated on the same held-out test set. The models fall into two groups: the statistical baselines and the MTL models (Sections~\ref{sec:stat-baselines} and~\ref{sec:mtl-baselines}) predict \verb|CHIRAL| as an additional target, whereas the remaining two models receive \verb|CHIRAL| as an input feature (Section~\ref{sec:chiral-input}). The BiLSTM model shares the final model's dual-branch architecture (Fig.~\ref{fig:architecture}), except that its time-series branch uses a bidirectional LSTM in place of the Transformer encoder; comparing the two therefore isolates the effect of the time-series encoder. The final Transformer attains the highest $R^2$ on every target and the highest $R^2$ averaged over the three targets ($0.458$). Relative to the BiLSTM, the largest gain is for \verb|GL_Orientation| ($0.311$ versus $0.177$), and the Transformer also attains the higher $R^2$ for every target within every severity bin of Table~\ref{tab:severity}. For both encoders, the model that receives \verb|CHIRAL| as an input attains a higher mean $R^2$ than its MTL counterpart ($0.360$ versus $0.329$ for the BiLSTM, and $0.458$ versus $0.406$ for the Transformer).

    \begin{table}[!t]
    \renewcommand{\arraystretch}{1.2}
    \caption{Test-Set $R^2$ of the Evaluated Architectures}
    \label{tab:comparison}
    \centering
    \setlength{\tabcolsep}{4pt}
    \begin{tabular}{lcccc}
    \hline
    Model & \texttt{TOPMORPH} & \texttt{Bpar} & \texttt{Orient.} & Mean \\
    \hline
    \multicolumn{5}{l}{\textit{\texttt{CHIRAL} as a target}} \\
    RF                  & 0.424 & 0.192 & 0.052 & 0.223 \\
    XGBoost             & 0.419 & 0.215 & 0.051 & 0.228 \\
    TSF                 & 0.164 & 0.071 & 0.020 & 0.085 \\
    BiLSTM-MTL          & 0.501 & 0.311 & 0.175 & 0.329 \\
    MTL-DNN             & 0.569 & 0.317 & 0.240 & 0.375 \\
    Transformer-MTL     & 0.562 & 0.377 & 0.277 & 0.406 \\
    \hline
    \multicolumn{5}{l}{\textit{\texttt{CHIRAL} as an input}} \\
    BiLSTM              & 0.563 & 0.341 & 0.177 & 0.360 \\
    Transformer (final) & \textbf{0.635} & \textbf{0.429} & \textbf{0.311} & \textbf{0.458} \\
    \hline
    \multicolumn{5}{p{0.93\columnwidth}}{\footnotesize
    Mean: $R^2$ averaged over the three targets. BiLSTM and Transformer: same architectures as BiLSTM-MTL and Transformer-MTL, without the \texttt{CHIRAL} output head and with \texttt{CHIRAL} as an input. Orient.:
    \texttt{GL\_Orientation}. Abbreviations as in Table~\ref{tab:chiral}.}
    \end{tabular}
    \end{table}

\subsection{Overall Performance}

    Table~\ref{tab:overall} reports the final model's performance on the held-out test set ($163{,}246$ rows from $4{,}521$ simulations). \verb|TOPMORPH| is recovered most accurately ($R^2 = 0.635$), followed by \verb|Bpar| ($R^2 = 0.429$) and \verb|GL_Orientation| ($R^2 = 0.311$). The rank-based Spearman correlations follow the same ordering, and the mean signed error is small for every target, at most $1.4\%$ of the target's standard deviation, so the predictions are not systematically biased over the test set as a whole. Prediction errors are only weakly correlated across targets: the Pearson correlation between the errors of \verb|TOPMORPH| and \verb|Bpar| is $0.165$, and the other two pairs have $|r| \le 0.026$.

    \begin{table}[!t]
    \renewcommand{\arraystretch}{1.2}
    \caption{Overall Test-Set Performance of the Final Model}
    \label{tab:overall}
    \centering
    \setlength{\tabcolsep}{3.5pt}
    \begin{tabular}{lccccc}
    \hline
    Target & $R^2$ & RMSE & $r$ & $\rho$ & Bias \\
    \hline
    \texttt{TOPMORPH}        & 0.635 & 0.669 & 0.797 & 0.763 & $-0.013$ \\
    \texttt{Bpar}$^{\ast}$   & 0.429 & 0.799 & 0.655 & 0.620 & $-0.004$ \\
    \texttt{GL\_Orientation} & 0.311 & 85.3  & 0.559 & 0.479 & $\phantom{-}1.46$ \\
    \hline
    \multicolumn{6}{p{0.93\columnwidth}}{\footnotesize
    $r$: Pearson correlation; $\rho$: Spearman correlation; Bias: mean signed
    error (predicted minus true). RMSE and bias are in the units of each target;
    $^{\ast}$\texttt{Bpar} RMSE and bias are given in units of $10^{-3}$.}
    \end{tabular}
    \end{table}

\subsection{Dependence on CME Severity (peak magnetic field $B_m$ strength)}

\begin{figure*}[!t]
    \centering
    \includegraphics[width=\textwidth]{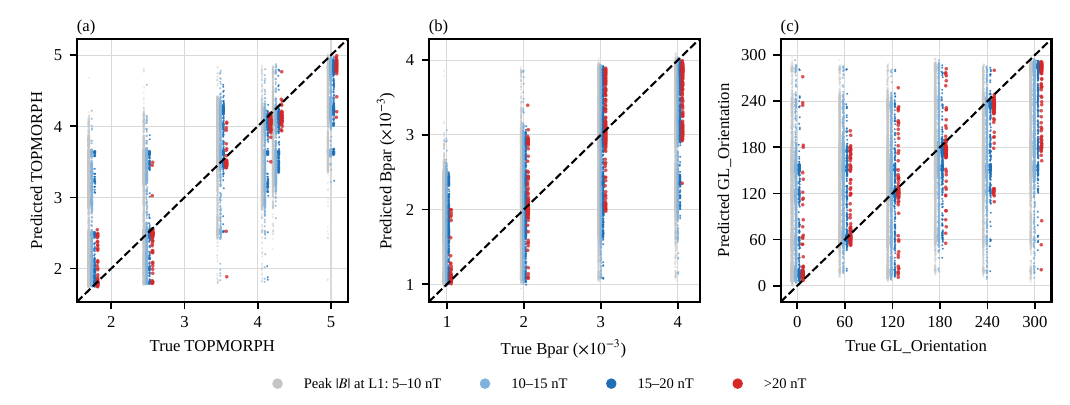}
    \caption{Predicted versus true values on the test set for (a) \texttt{TOPMORPH},
    (b) \texttt{Bpar}, and (c) \texttt{GL\_Orientation}, colored by CME severity
    (peak $|B|$ at L1). Points are offset horizontally by severity bin for visibility;
    all points in a group share the same true value. The dashed line marks perfect
    prediction.}
    \label{fig:pred-true}
    \end{figure*}

    Prediction accuracy increases monotonically with CME severity for all three targets (Table~\ref{tab:severity}). From the weakest ($5$--$10$~nT) to the strongest ($>20$~nT) bin, $R^2$ rises from $0.607$ to $0.967$ for \verb|TOPMORPH|, from $0.377$ to $0.860$ for \verb|Bpar|, and from $0.263$ to $0.833$ for \verb|GL_Orientation|. RMSE decreases correspondingly, from $0.699$ to $0.154$ for \verb|TOPMORPH|, from $0.831{\times}10^{-3}$ to $0.368{\times}10^{-3}$ for \verb|Bpar|, and from $87.9$ to $41.2$ for \verb|GL_Orientation|. Fig.~\ref{fig:pred-true} shows the same trend: predictions for weak events spread across nearly the full range of each target, whereas predictions for the strongest events concentrate near the true values. Prediction bias remains small in every bin: at most $1.7\%$ of the within-bin standard deviation for \verb|TOPMORPH| and \verb|GL_Orientation|, and at most $3.6\%$ for \verb|Bpar|, which is slightly under-predicted in the $>20$~nT bin. The $>20$~nT bin is also the smallest of the four, containing $1{,}175$ rows drawn from $319$ simulations. Conversely, the $5$--$10$~nT bin contains $75.2\%$ of all test rows and therefore dominates the overall metrics of Table~\ref{tab:overall}: restricted to rows with peak $|B| \ge 10$~nT, $R^2$ rises to $0.720$ for \verb|TOPMORPH|, $0.574$ for \verb|Bpar|, and $0.454$ for \verb|GL_Orientation|.

    \begin{table}[!t]
    \renewcommand{\arraystretch}{1.2}
    \caption{Test-Set $R^2$ and RMSE by CME Severity Bin}
    \label{tab:severity}
    \centering
    \setlength{\tabcolsep}{3pt}
    \begin{tabular}{lrrccc}
    \hline
    Peak $|B|$ & Rows & Sims. & \texttt{TOPMORPH} & \texttt{Bpar}$^{\ast}$ & \texttt{Orient.} \\
    \hline
    \multicolumn{6}{l}{\textit{$R^2$}} \\
    $5$--$10$ nT  & 122{,}830 & 4{,}521 & 0.607 & 0.377 & 0.263 \\
    $10$--$15$ nT &  32{,}940 & 2{,}394 & 0.706 & 0.547 & 0.426 \\
    $15$--$20$ nT &   6{,}289 & 1{,}062 & 0.765 & 0.627 & 0.531 \\
    $>20$ nT      &   1{,}175 &     319 & 0.967 & 0.860 & 0.833 \\
    \hline
    \multicolumn{6}{l}{\textit{RMSE}} \\
    $5$--$10$ nT  & & & 0.699 & 0.831 & 87.9 \\
    $10$--$15$ nT & & & 0.591 & 0.710 & 78.7 \\
    $15$--$20$ nT & & & 0.506 & 0.651 & 71.2 \\
    $>20$ nT      & & & 0.154 & 0.368 & 41.2 \\
    \hline
    \multicolumn{6}{p{0.93\columnwidth}}{\footnotesize
    Peak $|B|$: maximum magnetic field magnitude observed at L1. Sims.: number of
    distinct simulations contributing rows to the bin; a simulation can contribute
    rows to several bins. RMSE is in the units of each target;
    $^{\ast}$\texttt{Bpar} RMSE is given in units of $10^{-3}$. Orient.:
    \texttt{GL\_Orientation}. Overall values are given in Table~\ref{tab:overall}.}
    \end{tabular}
    \end{table}
    
\section{Conclusion}\label{sec:conclusions}
\label{sec:conclusion}

    This work posed the recovery of a CME's initial magnetic field-related Gibson-Low flux-rope parameters from its L1 time series as a supervised inverse problem, trained on a grid of GAMERA+Gibson-Low simulations. Among the architectures compared, a dual-branch Transformer that combines the raw L1 series with a scalar context vector attains the highest accuracy on all three continuous targets (mean $R^2 = 0.458$; Table~\ref{tab:comparison}). Chirality, in contrast, could not be predicted above the majority-class baseline by any architecture (Table~\ref{tab:chiral}), and for both the BiLSTM and Transformer encoders, the model that receives \verb|CHIRAL| as an input attains a higher mean $R^2$ than its multi-task counterpart. Prediction accuracy increases monotonically with CME severity: for CMEs with peak $|B| > 20$~nT, $R^2$ reaches $0.967$ for \verb|TOPMORPH|, $0.860$ for \verb|Bpar|, and $0.833$ for \verb|GL_Orientation|, so the model is most reliable for the events of greatest space weather impacts.

    The results demonstrate a principal possibility to speed up the direct modeling exercise for historically-observed CME events. Given that the accuracy increases towards the CMEs with the stronger magnetic field (here, in terms of the peak $B_m$ value), the recovery of the initial magnetic properties can be done more reliably for the strong CME events which are, in turn, most geoeffective and of the most interest from the operational standpoint. In the next steps, we plan to utilize the inverse model on $\sim$50 historic CME events observed during 2010-2026 and guide the corresponding parameter selection in GAMERA-GL simulations.

    The results also expose the limits of the present formulation. The inversion is ill-posed: different flux-rope configurations can produce similar signatures at L1, which is consistent with the chance-level chirality results and with the wide spread of predictions for weak events (Fig.~\ref{fig:pred-true}). A deterministic regressor returns a single estimate in such cases and gives no indication of which predictions are ambiguous. In addition, all three targets are trained with a squared-error loss, although \verb|TOPMORPH| is sampled at only six values in the simulation grid, \verb|Bpar| has been initialized only with four options, and the flux-rope orientation is an angular quantity with six options (see Figure~\ref{fig:cmeparameters}). The further assimilation of upcoming GAMERA-GL direct modeling results will help to enhance the robustness of the model.

    Future work will therefore reformulate the problem as posterior estimation of the GL flux-rope parameters given the L1 observations, which would quantify the uncertainty of each predicted parameter. The potential way to do it is to use conditional generative models such as conditional variational autoencoders \cite{Sohn:2015:CVAE}, invertible neural networks \cite{Ardizzone:2019:INN}, or simulation-based inference frameworks such as BayesFlow \cite{Radev:2022:BayesFlow}, with output distributions matched to each parameter. Further directions include training strategies that emphasize strong events (which, effectively, for a small minority class in the sampled dataset), additional information that may constrain chirality (such as the magnetic configuration of the source active region), ablation study for the kinematic CME properties and boundary conditions (to quantify how their precision affects the results), and the application of the trained models to historical CME events to provide initial parameter estimates for forward modeling. In addition, the background subtraction and isolation of the CME impact may be helpful for pruning the considered dataset further and avoiding utilization of background solar wind-only samples in training.

\section*{Acknowledgment}

We would like to thank the NASA LWS project team members for their support and the availability of the modeling data. We also thank Dr.~Hemanth Demakethepalli Venkateswara for helpful discussions on the probabilistic formulation of the inverse problem, which informed the future directions outlined in Section~\ref{sec:conclusion}. This project has been supported in part by funding from NASA's Living With a Star program grant \#80NSSC24K1111. We acknowledge NASA HECC for the use of computational resources. Additionally, we acknowledge the use of the PROFLEX-CI computing resources at Georgia State University, which are supported by the U.S. National Science Foundation under Grant No. 2430193.

\bibliographystyle{IEEEtran}
\bibliography{cmeinverse}

\end{document}